\documentclass[twocolumn]{aastex701}
\usepackage{siunitx}
\usepackage{xfrac}

\DeclareSIUnit{\cMpc}{cMpc}

\usepackage{xcolor}
\usepackage[normalem]{ulem}

\newcommand{\rem}[1]{
  \bgroup
  \markoverwith{\textcolor{blue}{\rule[0.5ex]{2pt}{0.4pt}}}
  \ULon{#1}
  \egroup
}

\def\ep{{EP}}
\def\fxt{{FXT}}
\def\epfxt{{EP-FXT}}
\def\xmm{\textit{XMM-Newton}}

\def\erosita{{eROSITA}}

\def\nustar{\textit{NuSTAR}}

\def\ha{\ensuremath{\mathrm{H}\alpha}}
\def\hb{\ensuremath{\mathrm{H}\beta}}
\def\hg{\ensuremath{\mathrm{H}\gamma}}

\def\oiii{\ensuremath{[\mathrm{O}\,\mathrm{III}]}}
\def\nii{\ensuremath{[\mathrm{N}\,\mathrm{II}]}}
\def\sii{\ensuremath{[\mathrm{S}\,\mathrm{II}]}}

\def\efxtdas{\textsc{eFXTDAS}}
\def\fxtdas{\textsc{FXTDAS}}
\def\xstack{\textsc{Xstack}}
\def\topcat{\textsc{TOPCAT}}
\def\ppxf{\textsc{pPXF}}
\def\xspec{\textsc{Xspec}}

\def\mekal{\texttt{mekal}}

\begin{document}

\title{Local Analogs of Little Red Dots: Clues to the X-ray Weakness}

\author[0009-0004-9950-9807]{Shi-Jiang Chen}
\affiliation{Department of Astronomy, University of Science and Technology of China, Hefei 230026, People's Republic of China}
\affiliation{School of Astronomy and Space Science, University of Science and Technology of China, Hefei 230026, People's Republic of China}
\email{csj666@mail.ustc.edu.cn}

\author[0000-0002-4419-6434]{Jun-Xian Wang}
\affiliation{Department of Astronomy, University of Science and Technology of China, Hefei 230026, People's Republic of China}
\affiliation{School of Astronomy and Space Science, University of Science and Technology of China, Hefei 230026, People's Republic of China}
\affiliation{College of Physics, Guizhou University, Guiyang, Guizhou, 550025, People's Republic of China}
\email{jxw@ustc.edu.cn}

\author[0000-0002-9634-2923]{Zhen-Ya Zheng}
\affiliation{Shanghai Astronomical Observatory, Chinese Academy of Sciences, 80 Nandan Road, Shanghai 200030, People's Republic of China}
\email{zhengzy@shao.ac.cn}

\author[0000-0003-3987-0858]{Ruqiu Lin}
\affiliation{Shanghai Astronomical Observatory, Chinese Academy of Sciences, 80 Nandan Road, Shanghai 200030, People's Republic of China}
\affiliation{University of Massachusetts Amherst, 710 North Pleasant Street, Amherst, MA 01003-9305, USA}
\email{ruqiulin@umass.edu}

\author[0000-0001-6947-5846]{Luis C. Ho}
\affiliation{Kavli Institute for Astronomy and Astrophysics, Peking University, Beijing 100871, China}
\affiliation{Department of Astronomy, School of Physics, Peking University, Beijing 100871, China}
\email{lho.pku@gmail.com}

\author[0000-0002-4223-2198]{Zhen-Yi Cai}
\affiliation{Department of Astronomy, University of Science and Technology of China, Hefei 230026, People's Republic of China}
\affiliation{School of Astronomy and Space Science, University of Science and Technology of China, Hefei 230026, People's Republic of China}
\email{zcai@ustc.edu.cn}

\author[0009-0003-4721-177X]{Chang-Hao Chen}
\affiliation{Kavli Institute for Astronomy and Astrophysics, Peking University, Beijing 100871, China}
\affiliation{Department of Astronomy, School of Physics, Peking University, Beijing 100871, China}
\email{cchen_louis@stu.pku.edu.cn}

\author[0000-0001-9274-1145]{Zsofi Igo}
\affiliation{European Space Agency (ESA), European Space Astronomy Centre (ESAC), Camino Bajo del Castillo s/n, E-28692 Villanueva de la Ca\~{n}ada, Madrid, Spain}
\email{zsofi.igo@esa.int}

\author[0000-0003-2280-2904]{Jia-Lai Kang}
\affiliation{Department of Astronomy, University of Science and Technology of China, Hefei 230026, People's Republic of China}
\affiliation{School of Astronomy and Space Science, University of Science and Technology of China, Hefei 230026, People's Republic of China}
\email{ericofk@mail.ustc.edu.cn}

\author[0000-0001-8496-4162]{Ruancun Li}
\affiliation{Max-Planck-Institut für Extraterrestrische Physik (MPE), Giessenbachstrasse 1, 85748 Garching bei München, Germany}
\email{liruancun@mpe.mpg.de}

\author[0000-0002-2941-6734]{Teng Liu}
\affiliation{Department of Astronomy, University of Science and Technology of China, Hefei 230026, People's Republic of China}
\affiliation{School of Astronomy and Space Science, University of Science and Technology of China, Hefei 230026, People's Republic of China}
\email{liuteng@ustc.edu.cn}

\author[0000-0002-3742-6609]{Wenke Ren}
\affiliation{Shanghai Astronomical Observatory, Chinese Academy of Sciences, 80 Nandan Road, Shanghai 200030, People's Republic of China}
\email{wenke.ren@shao.ac.cn}

\author[0000-0001-8515-7338]{Zhen-Bo Su}
\affiliation{Department of Astronomy, University of Science and Technology of China, Hefei 230026, People's Republic of China}
\affiliation{School of Astronomy and Space Science, University of Science and Technology of China, Hefei 230026, People's Republic of China}
\email{zbsu@mail.ustc.edu.cn}

\correspondingauthor{\\ Shi-Jiang Chen, Jun-Xian Wang, \\ Zhen-Ya Zheng, Ruqiu Lin}
\email{csj666@mail.ustc.edu.cn, jxw@ustc.edu.cn, \\ zhengzy@shao.ac.cn, ruqiulin@umass.edu}

\begin{abstract}

A major question about JWST's little red dots (LRDs) at $z\sim4-8$ is the origin of their apparent X-ray weakness. Even deep Chandra stacks yield only tentative detections or broad-band limits and cannot distinguish among super-Eddington accretion, intrinsic X-ray weakness, and Compton-thick (CT) obscuration.
Local analogs offer a powerful complementary route. At $z\sim0.3$, similarly luminous sources require $\sim10^{3-4}$ less exposure for comparable signal-to-noise, suffer less source confusion, and directly probe the diagnostic rest-frame soft X-rays.
Here we present a pilot Einstein Probe/FXT study of six ``local little red dots'' (LLRDs), V-shaped broad-line green pea galaxies selected as low-redshift counterparts of LRDs.
These sources show X-ray diversity: ``X-normal'' LLRDs resemble type 1 AGNs in X-ray luminosity and optical variability, whereas ``X-weak'' stack lies $\sim$\qtyrange[range-phrase=--,range-units=single]{0.5}{0.7}{dex} below standard unobscured-quasar scaling relations while remaining compatible with the representative LRD stack and $\qty{95}{\%}$ of published individual limits. The X-weak LLRDs also lack detected optical variability and have redder optical continua than X-normal LLRDs, making their V-shaped SEDs closer to those of high-redshift LRDs.
The detection of the X-weak stack at $\sim3.3\sigma$ with the soft-X-ray-sensitive EP-FXT provides rest-frame $\sim2-3\,\mathrm{keV}$ constraints difficult to reconcile with a typical unobscured super-Eddington model or a local-AGN-like CT geometry examined here, although modified variants remain possible. The simplest alternative is a mildly X-ray-weak corona ($\Delta\alpha_\mathrm{ox}\sim-0.1$ to $-0.2$) with Compton-thin absorption ($N_\mathrm{H}\sim10^{22-23}\,\mathrm{cm}^{-2}$), consistent with the independent Balmer decrement/break diagnostics.
Together with growing samples of local LRD analogs, our results offer a practical path for testing the physical nature of LRD X-ray weakness.

\end{abstract}



\section{The LRD X-ray bottleneck}\label{sec:intro}

JWST has revealed an abundant population of enigmatic galaxies at $z\sim4-8$ that are extremely compact and exhibit characteristic ``V-shaped'' rest-frame UV-to-optical spectral energy distributions (SEDs), now commonly referred to as ``little red dots'' (LRDs; \citealt{Harikane2023Jwstnirspec,Matthee2024Little,Greene2024Uncover,Kocevski2025Rise,Labbe2025Uncover,Zhang2026Composite}). Broad permitted emission lines with widths of $\gtrsim10^3\,\mathrm{km\,s^{-1}}$ are found in a substantial fraction of LRDs, strongly suggesting accretion onto massive BHs. Nevertheless, the conventional multiwavelength signatures of AGNs are often weak or absent \citep[e.g.,][]{Akins2025Cosmosweb,Wang2026Missing}. In particular, LRDs are rarely detected individually in X-rays \citep[e.g.,][]{Ananna2024Xray}, apart from a small number of exceptions \citep[e.g.,][]{Fu2026Discovery,Hviding2026Xray}. Stacking analyses totaling tens to hundreds of megaseconds have yielded only tentative signals or nondetections. When compared with the X-ray luminosities expected from standard AGN relations, these measurements indicate that LRDs are X-ray faint on average \citep{Yue2024Stacking,Sacchi2025Chandra}. Similar X-ray weakness has also been reported among the broader population of JWST-selected high-redshift AGNs \citep[e.g.,][]{Maiolino2025Jwst,Brazzini2026Little,Comastri2026Jwstdiscovered}, possibly suggesting a common physical condition in rapidly growing BHs in the early universe.

Several scenarios have been proposed to explain this apparent population-level X-ray deficit. LRDs may host intrinsically weak or unusually soft X-ray coronae. In particular, super-Eddington accretion can produce steep X-ray spectra \citep[photon index $\Gamma\sim3$--$4$;][]{Pacucci2024Mildly,Inayoshi2025Weakness,Zhou2026Little,Lambrides2026Case}, leaving few detectable photons in the rest-frame hard X-rays sampled by Chandra at $z\sim4$--$8$. Rapid accretion would also help explain how BHs reach inferred masses of $\sim10^{7-8}\,M_\odot$ so early in cosmic history without requiring extremely massive seeds \citep[e.g.,][]{DeLuca2020Evolution,Yuan2024Rapidly}. Alternatively, absorption by dense nuclear gas may suppress the observed X-ray emission. Strong evidence for such gas comes from the combined presence of Balmer absorption, prominent Balmer breaks, red optical continua, and elevated broad-line Balmer decrement---the ``ABCD'' diagnostics (\citealt{Chen2026Abcd,Inayoshi2025Critical}; see also \citealt{Juodbalis2024Jades,Ji2025Blackthunder,Naidu2025Black,Chang2026Impact,Zhuang2026Nexus,Matthee2026Engine,Juodbalis2026Incomplete}). Together, these features can be explained self-consistently by a dense, optically thick nuclear medium with $n_\mathrm{H}\gtrsim10^9\,\mathrm{cm}^{-3},\ N_\mathrm{H}\gtrsim10^{22}\,\mathrm{cm}^{-2}$ and substantial covering factor in at least some systems. Such an environment makes X-ray obscuration a natural possibility, and extreme CT columns of $N_\mathrm{H}\gtrsim10^{24.5}\,\mathrm{cm}^{-2}$ \citep[e.g.,][]{Maiolino2025Jwst,Rusakov2026Little,Sneppen2026Inside} have been proposed to explain the reported LRD X-ray deficit.

Testing these scenarios requires establishing both the magnitude and prevalence of the X-ray deficit. Yet even megasecond Chandra observations have so far provided weak individual constraints: only about half of broad-line LRDs have upper limits below the X-ray luminosities predicted by standard AGN relations \citep{Yue2024Stacking,Casey2026Population}. Stacking provides stronger population-level constraints, but low-count stacks of heterogeneous samples spanning broad ranges in redshift and luminosity generally require careful interpretation. Moreover, distinguishing the physical origin of any deficit remains difficult because Chandra observations at $z\sim6$ predominately sample rest-frame hard energies, whereas the soft X-ray in which steep intrinsic spectra and heavy obscuration differ most strongly lies below its effective bandpass.

Nearby analogs offer a direct route around this observational bottleneck. For sources of comparable intrinsic luminosity, reaching similar X-ray constraints at $z\sim0.3$ requires $\sim10^{3-4}$ less exposure than at $z\sim6$, depending on the source-to-background ratio. Their proximity also reduces physical blending at a given angular resolution and, most importantly, brings the diagnostic rest-frame soft X-ray emission into the observable band.

Can we find local analogs? Well before the launch of JWST, nearby dwarf galaxies were already studied as local analogs of early galaxies \citep[e.g.,][]{Simmonds2016Some,Mezcua2017Observational}. In particular, green pea galaxies (GPs), a class of compact dwarf galaxies with strong \oiii~emission, resemble high-redshift star-forming galaxies in stellar mass, size, specific star-formation rate, metallicity, and, in some cases, Ly$\alpha$ and LyC escape \citep{Yang2016Green,Izotov2016Eight}. From an initial sample of 2190 GPs, \citet{Lin2024Intermediatemass} identified 59 candidates with broad emission lines indicative of AGN activity. Building on this sample, \citet{Lin2025Discovery} identified seven V-shaped broad-line GPs whose UV-to-optical SEDs and compact morphologies (unresolved or $\lesssim1\,\mathrm{kpc}$ at DESI-LS10 resolution\footnote{At this resolution, their optical profiles are consistent with, or more compact than, expected from the star-forming-galaxy mass–size relation \citep{Nedkova2021Extending}.}) satisfy commonly used observational LRD criteria \citep[e.g.,][]{Labbe2025Uncover,Kocevski2025Rise,Hviding2025Rubies}. They also resemble LRDs in their ionization conditions and apparently overmassive BHs. We therefore refer to these systems as ``local little red dots'' (LLRDs), without assuming a common physical origin for all LRDs. Demographically, broad-line and V-shaped broad-line systems account for 2.7\% and 0.32\%, respectively, of the parent GP sample. These fractions are interestingly of the same order as the ratio of the number density of LRDs at $z\sim$ \numrange[range-phrase=--]{6}{7} ($\sim10^{-5}$--$10^{-4}\,\mathrm{cMpc}^{-3}$, \citealt{Pan2026Nexus} and references therein) to that of galaxies with $8.0<\log (M_\star/M_\odot)<9.5$ at similar redshifts ($\sim$\qtyrange[range-phrase=--,range-units=single]{3e-3}{4e-3}{\per\cubic\cMpc}, \citealt{Weibel2024Galaxy}), or roughly 0.25--3\%. Complementary searches have subsequently enlarged the population of candidate low-redshift LRD analogs using alternative methods \citep[e.g.,][]{Lin2026Discovery,Lin2026Lrds,Ji2026Lord,Ding2026Discovery,Rinaldi2026Beyond}.

Here we present a pilot X-ray study of six of the seven LLRDs identified by \citet{Lin2025Discovery}, using pointed Einstein Probe \citep[\ep,][]{Yuan2025Science} \fxt~observations obtained through our program (PI: J.-X. Wang). The remaining LLRD is excluded due to strong X-ray contamination by a DESI broad-line AGN $\sim\ang{;;20}$ away. Even before considering their X-ray emission, these six LLRDs show internal diversity: J0928 and J1154 are the only two with securely detected AGN-like optical variability and also exhibit bluer optical continua than the other four objects \citep{Lin2025Discovery,Lin2026Local}. Although all six satisfy the original LLRD selection, these independent optical differences provide important context for interpreting their X-ray properties. We therefore ask whether this optical diversity is accompanied by a corresponding contrast in X-ray emission and what that contrast may reveal about high-redshift LRDs. Our observational results are presented in Section~\ref{sec:results}, followed by their implications for high-redshift LRDs in Section~\ref{sec:discussion}. Throughout this work, we adopt a cosmology with $H_0=\qty{67.7}{km.s^{-1}.Mpc^{-1}}$, $\Omega_{\mathrm{m}}=0.31$, and $\Omega_\Lambda=0.69$ \citep{Planck2020Planck}.

\section{X-ray reveals two types of LLRDs}\label{sec:results}
\subsection{Direct view from \epfxt~X-ray images}
\begin{figure*}
    \centering
    \includegraphics[width=0.9\linewidth]{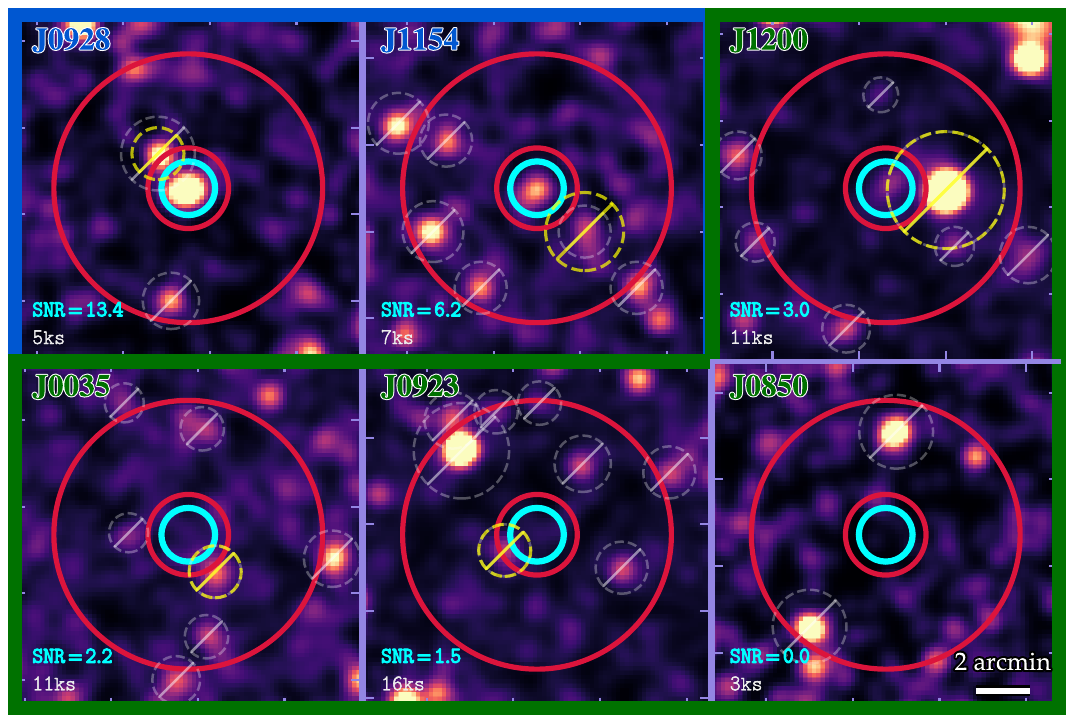}
    \caption{Observed-frame \epfxt~\qtyrange[range-phrase=--,range-units=single]{0.5}{4}{keV} exposure-corrected count-rate images, coadded over all observations and detector modules and smoothed with a Gaussian kernel of $\sigma=2$ pixels ($\simeq\ang{;;19}$). Cyan circles and crimson annuli mark the source and background apertures; yellow and gray circles show contaminant masks. Labels give the extraction-based SNR and effective exposure. Blue and green frames denote X-normal and X-weak LLRDs.}
    \label{fig:xray_image}
\end{figure*}

We reduce the \epfxt~data summarized in Appendix \ref{app:obs_summary} with our custom \efxtdas\footnote{\url{https://github.com/AstroChensj/eFXTDAS}} tool, which extends the official \fxtdas\footnote{\url{https://epfxt.ihep.ac.cn/analysis}} pipeline (Appendix~\ref{app:efxtdas}). Following calibration and event cleaning, we coadd the observed-frame \qtyrange[range-phrase=--,range-units=single]{0.5}{4}{keV} images from all observations and both modules and perform wavelet- and PSF-based source detection. Spectra are extracted within $\ang{;;60}$ (enclosing $\sim\qty{90}{\%}$ of the on-axis PSF energy), with local backgrounds from $\ang{;;90}$--$\ang{;;300}$ annuli. Contaminating sources are masked and the responses account for the resulting geometry and vignetting. From the extracted spectra, we calculate the background-only probability $p_\mathrm{null}$ and quote its one-sided Gaussian-equivalent significance as the source SNR (Appendix \ref{app:detection_significance}). 

As can be seen from the coadded images in Figure \ref{fig:xray_image}, X-ray reveals a contrast aligned with the pre-existing optical differences: J0928 and J1154 are significantly detected in X-rays, whereas J1200, J0035, J0923, and J0850 are undetected or only marginally detected despite comparable or longer exposures. We hereafter refer to these two subsets as ``X-normal'' and ``X-weak'' LLRDs, respectively.

\subsection{Benchmarking LLRD X-ray emission against local AGNs and LRDs}\label{subsec:compare}
\begin{figure*}
    \centering
    \includegraphics[width=1.0\linewidth]{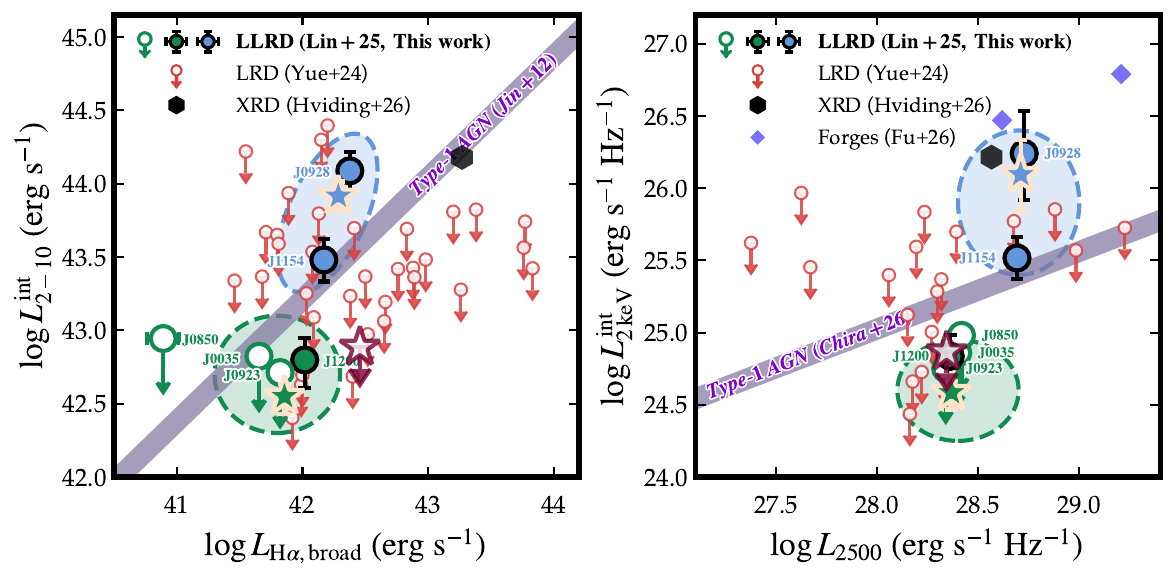}
    \caption{Intrinsic X-ray emission of LLRDs benchmarked against AGNs and high-redshift LRDs. Left: \qtyrange[range-phrase=--,range-units=single]{2}{10}{keV} luminosity versus broad-\ha~luminosity; right: \qty{2}{keV} monochromatic luminosity versus rest-frame \qty{2500}{\AA} luminosity. X-weak LLRDs are green (open circles: upper limits; filled circle: marginal J1200 detection), and X-normal LLRDs are blue; stars mark their spectral stacks, with dashed ellipses associating each stack with its members. Red circles and dark-red stars denote the individual and representative stacked $3\sigma$ LRD limits based on \citet{Yue2024Stacking}; the stacked limits use exposure-weighted count-rate conversions and geometric-mean horizontal coordinates. Black and purple symbols show the XRD and Forges. Purple lines show the type 1 AGN relations.}
    \label{fig:lx_lha_luv}
\end{figure*}
\begin{deluxetable*}{cccccccccccccc}
\tabletypesize{\tiny}
\setlength{\tabcolsep}{2pt}
\tablecaption{Properties of LLRDs and their stacked X-ray spectra\label{tab:lgd_properties}}
\tablehead{
\colhead{Name} &
\colhead{$z_{\rm spec}$} &
\colhead{$\log M_{\rm BH}$} &
\colhead{$\log M_\star$} &
\colhead{$\log {\rm SFR}$} &
\colhead{$SF_{\infty,\rm DRW}$} &
\colhead{$\log L_{{\rm H}\alpha,{\rm broad}}$} &
\colhead{$\log L_{2500}$} &
\colhead{SNR} &
\colhead{$\log L_{2-10}$} &
\colhead{$\log L_{\rm 2\,keV}$} &
\colhead{$N_{\rm H}$} &
\colhead{$\Gamma$} &
\colhead{$C_f$} \\
\colhead{} &
\colhead{} &
\colhead{($M_\odot$)} &
\colhead{($M_\odot$)} &
\colhead{($M_\odot\,{\rm yr}^{-1}$)} &
\colhead{} &
\colhead{(erg s$^{-1}$)} &
\colhead{(erg s$^{-1}$ Hz$^{-1}$)} &
\colhead{} &
\colhead{(erg s$^{-1}$)} &
\colhead{(erg s$^{-1}$ Hz$^{-1}$)} &
\colhead{($10^{22}$ cm$^{-2}$)} &
\colhead{} &
\colhead{}
}
\startdata
\cutinhead{Individual sources}
J0928 & 0.293 & $7.82\pm0.01$ & $10.02\pm0.04$ & $0.65\pm0.01$ & $0.18_{-0.02}^{+0.03}$ & $42.38\pm0.01$ & $28.73$ & 13.4 & $44.09_{-0.13}^{+0.13}$ & $26.24_{-0.32}^{+0.29}$ & $9.01_{-3.27}^{+3.20}$ & $2.13_{-0.71}^{+0.67}$ & $0.94_{-0.13}^{+0.04}$ \\
J1154 & 0.269 & $7.68\pm0.03$ & $10.08\pm0.10$ & $0.26\pm0.02$ & $0.18_{-0.03}^{+0.06}$ & $42.17\pm0.02$ & $28.69$ & 6.2 & $43.48_{-0.15}^{+0.14}$ & $25.52_{-0.15}^{+0.14}$ & $3.61_{-1.19}^{+1.65}$ & $1.80(*)$ & $1.00(*)$ \\
J1200 & 0.303 & $8.47\pm0.03$ & $9.90\pm0.03$ & $0.74\pm0.01$ & -- & $42.02\pm0.02$ & $28.36$ & 3.0 & $42.80_{-0.19}^{+0.15}$ & $24.84_{-0.19}^{+0.15}$ & $1.00(*)$ & $1.80(*)$ & $1.00(*)$ \\
J0035 & 0.242 & $6.60\pm0.12$ & $8.71\pm0.03$ & $0.94\pm0.01$ & -- & $41.65\pm0.08$ & $28.40$ & 2.2 & $<42.82$ & $<24.86$ & $1.00(*)$ & $1.80(*)$ & $1.00(*)$ \\
J0923 & 0.323 & $6.92\pm0.04$ & $10.03\pm0.20$ & $0.52\pm0.01$ & -- & $41.82\pm0.03$ & $28.34$ & 1.5 & $<42.71$ & $<24.75$ & $1.00(*)$ & $1.80(*)$ & $1.00(*)$ \\
J0850 & 0.280 & $6.33\pm0.19$ & $8.33\pm0.08$ & $0.38\pm0.01$ & -- & $40.89\pm0.13$ & $28.42$ & 0.0 & $<42.94$ & $<24.98$ & $1.00(*)$ & $1.80(*)$ & $1.00(*)$ \\
\cutinhead{Stacked spectra}
J0928+J1154 & -- & $7.76\pm0.01$ & $10.05\pm0.06$ & $0.50\pm0.01$ & -- & $42.29\pm0.01$ & $28.71$ & 14.8 & $43.92_{-0.10}^{+0.10}$ & $26.10_{-0.26}^{+0.24}$ & $9.34_{-2.59}^{+2.37}$ & $2.22_{-0.57}^{+0.57}$ & $0.95_{-0.06}^{+0.03}$ \\
J1200+J0923+J0035 & -- & $8.01\pm0.03$ & $9.81\pm0.11$ & $0.77\pm0.01$ & -- & $41.86\pm0.02$ & $28.37$ & 3.5 & $42.55_{-0.12}^{+0.08}$ & $24.59_{-0.12}^{+0.08}$ & $1.00(*)$ & $1.80(*)$ & $1.00(*)$ \\
\enddata
\tablecomments{Black-hole and stellar masses and broad \ha~luminosities are from \citet{Lin2025Discovery}, and optical-variability measurements are from \citet{Lin2026Local}. SFRs are derived from $\log\mathrm{SFR}=\log L_{\mathrm{H}\alpha,\mathrm{narrow}}-41.27$ following \citet{Kennicutt2012Star}. $L_{2500}$ is the interpolated, Galactic-extinction-corrected rest-frame \qty{2500}{\AA} monochromatic luminosity. X-ray SNRs are evaluated over the observed-frame \qtyrange[range-phrase=--,range-units=single]{0.5}{8}{keV} band for individual sources and the rest-frame \qtyrange[range-phrase=--,range-units=single]{0.5}{8}{keV} band for the spectral stacks. X-ray luminosities (\qty{2}{keV} monochromatic luminosity, \qtyrange[range-phrase=--,range-units=single]{2}{10}{keV} model luminosity) are absorption-corrected AGN power-law values, and $N_\mathrm{H}$ is the intrinsic nuclear column density. Parenthesized asterisks denote fixed parameters.}
\end{deluxetable*}

To benchmark LLRDs against AGNs and high-redshift LRDs, we fit their spectra with \xspec~\citep{Arnaud1996Xspec} using an absorbed AGN power-law plus fixed host-galaxy emission, attenuated by Galactic absorption: \texttt{TBabs*(zTBabs}\textsubscript{\texttt{agn}}\texttt{*zpowerlw}\textsubscript{\texttt{agn}}\texttt{+host)}. For X-weak LLRDs, we adopt $N_\mathrm{H}=10^{22}\,\mathrm{cm}^{-2}$ and $\Gamma=1.8$, consistent with \citet{Yue2024Stacking}, and fit only the normalization. For the X-normal LLRDs, their higher SNR allows $N_\mathrm{H}$ and, where warranted, the photon index to vary; J0928 additionally requires partial covering. 
Table \ref{tab:lgd_properties} reports the fitted parameters; Appendix \ref{app:Xnormal} presents details and the X-normal spectra (Figure \ref{fig:xray_spec}).
The host component includes X-ray binaries and hot gas normalized from the stellar mass and narrow-\ha-based SFR \citep{Lehmer2010Chandra,Mineo2012Xray,Kennicutt2012Star,Liu2024Origin}. Its luminosity is only $\sim10^{40}\,\mathrm{erg}\,\mathrm{s}^{-1}$, and even a dex uncertainty changes the reported nuclear luminosities or upper limits by $<5\%$.

To enhance SNR, we stack the rest-frame spectra of three X-weak LLRDs (J1200, J0035, and J0923) using \xstack~\citep{Chen2025Average}, which shifts each spectrum and corresponding Galactic-$N_\mathrm{H}$-corrected response to a common rest frame before combining them using average-luminosity response weighting (\texttt{LMN} mode). Here X-ray spectral stacking should be justified because the three objects span narrow ranges in redshift and broad-\ha~luminosity. J0850 is excluded because its broad-\ha~luminosity is $\sim\qty{1}{dex}$ lower and its shorter exposure would contribute mostly background. For comparison, the two X-normal LLRDs are stacked identically. Both results are reported in Table \ref{tab:lgd_properties}, with the X-weak stack reaching an SNR of $\sim3.5$ in the rest-frame \qtyrange[range-phrase=--,range-units=single]{0.5}{8}{keV} band.

Consistent with the direct view from Figure \ref{fig:xray_image}, the X-weak LLRD stack is $\sim$ \qtyrange[range-phrase=--,range-units=single]{1.4}{1.5}{dex} fainter in X-rays than the X-normal stack, although their \qty{2500}{\AA} and broad-\ha~luminosities differ by only $\sim$ \qty{0.4}{dex} (Figure \ref{fig:lx_lha_luv}). The same contrast is seen relative to the unobscured type 1 AGN relations from \citet{Jin2012CombinedIIRelation} and \citet{Chira2026Revisiting}: the marginal detection or upper limits of the individual X-weak LLRDs lie below the corresponding relations, except for J0850, whose shorter \epfxt~exposure does not provide a useful constraint in the broad-\ha~comparison. The X-weak stack lies $\sim\qty{0.55}{dex}$ below the \ha-based relation and $\sim\qty{0.65}{dex}$ below the \qty{2500}{\AA}-based relation. By contrast, the X-normal LLRDs are consistent with or brighter than the AGN relations, with their stack lying $\sim\qty{0.6}{dex}$ above them. As a descriptive check incorporating the three individual upper limits, we also report a reverse Kaplan–Meier estimate of the six-source X-ray luminosity distribution in Appendix \ref{app:km}.

To compare with high-redshift LRDs, Figure \ref{fig:lx_lha_luv} includes the individual LRD upper limits and a representative whole-sample stacked upper limit derived from \citet{Yue2024Stacking}. The X-weak LLRD stack lies below $\qty{95}{\%}$ ($19/20$) of the individual limits with available $L_{2500}$ measurements and remains compatible with the representative LRD stack. Thus, at the distance of the LLRDs, \epfxt~detects an X-ray luminosity that would remain below nearly all published individual high-redshift LRD limits, demonstrating the observational advantage of nearby analogs. For broader context, Figure \ref{fig:lx_lha_luv} also shows the ``X-ray dot'' (XRD) and ``Forges'', proposed transitional objects linking LRDs to more conventional quasars \citep{Hviding2026Xray,Fu2026Discovery}.

Importantly, the difference between the X-normal and X-weak LLRDs extends beyond X-rays. The two X-normal LLRDs are also the only objects with securely detected AGN-like optical variability \citep{Lin2026Local}, while X-weak LLRDs exhibit redder optical continua (Figure \ref{fig:sed}), larger mean offset from the local $M_\mathrm{BH}$--$M_\star$ relation (Table \ref{tab:lgd_properties}), and lack significant optical variability similar to the majority of LRDs \citep{Zhang2025Analysis}. Their Balmer properties further reinforce this contrast (Figure \ref{fig:bdbb}; Appendix \ref{app:balmer_decrement}). Joint fits to \ha, \hb, and \hg~show that the three X-weak LLRDs in the fiducial stack have a characteristic $\left(\mathrm{H}\alpha/\mathrm{H}\beta\right)_\mathrm{broad}\sim$\numrange[range-phrase=--]{7}{8} or larger and an average Balmer-break strength of $\sim1.6$. By contrast, X-normal LLRDs have $\left(\mathrm{H}\alpha/\mathrm{H}\beta\right)_\mathrm{broad}\sim3.5$, closer to Case B recombination (2.86) and typical type 1 AGN (3.1), together with a Balmer-break strength of $\sim1.1$.
These coherent multiwavelength differences motivate treating the two groups separately, with the X-weak LLRDs providing the closer local analogs for investigating the X-ray nature of high-redshift LRDs.

\section{Implications for high-redshift LRDs}\label{sec:discussion}
\begin{figure*}
    \centering
    \includegraphics[width=1.0\linewidth]{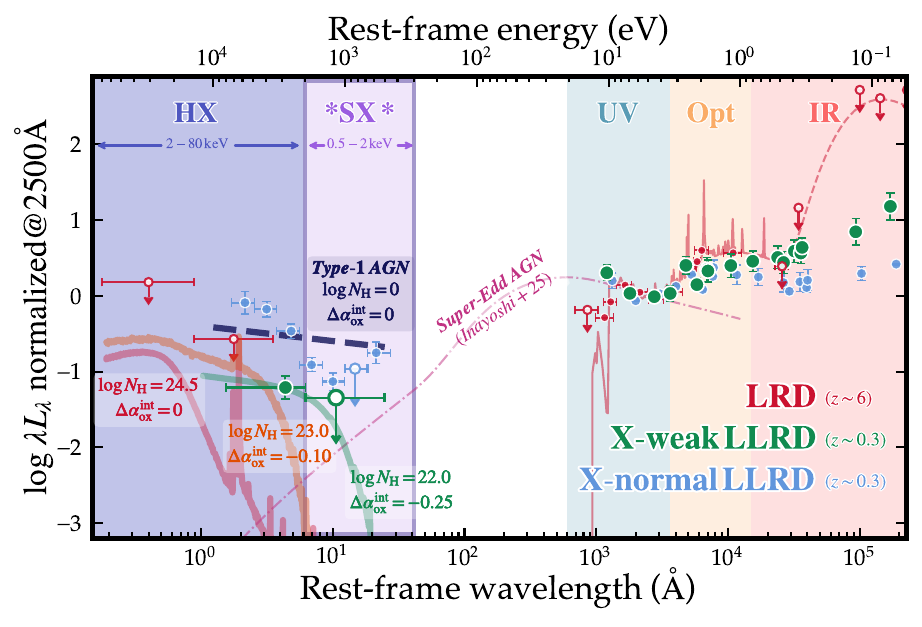}
    \caption{Normalized IR--optical--UV--X-ray SEDs of the X-weak LLRDs (green), X-normal LLRDs (blue), and high-redshift LRDs (red). All SEDs are normalized at the rest-frame \qty{2500}{\AA}. The LLRD measurements combine \epfxt~X-ray spectroscopy with GALEX, SDSS, SPHEREx, and WISE photometry. The red measurements and background SED model show the stacked LRD constraints from \citet{Akins2025Cosmosweb}. The curves show a typical unobscured super-Eddington model with $\dot m=1$ and $M_{\mathrm{BH}}=10^7\,M_\odot$ \citep{Inayoshi2025Weakness}; a CT model with $N_{\mathrm{H}}=10^{24.5}\,\mathrm{cm}^{-2}$ and $\Delta\alpha_{\mathrm{ox}}=0$; a Compton-thin model with $N_{\mathrm{H}}=10^{23}\,\mathrm{cm}^{-2}$ and $\Delta\alpha_{\mathrm{ox}}=-0.10$; the fitted X-weak LLRD model with $N_{\mathrm{H}}=10^{22}\,\mathrm{cm}^{-2}$ and $\Delta\alpha_{\mathrm{ox}}=-0.25$; and an unabsorbed local-AGN model with $\Delta\alpha_{\mathrm{ox}}=0$.}
    \label{fig:sed}
\end{figure*}
\begin{figure*}
    \centering
    \includegraphics[width=1.0\linewidth]{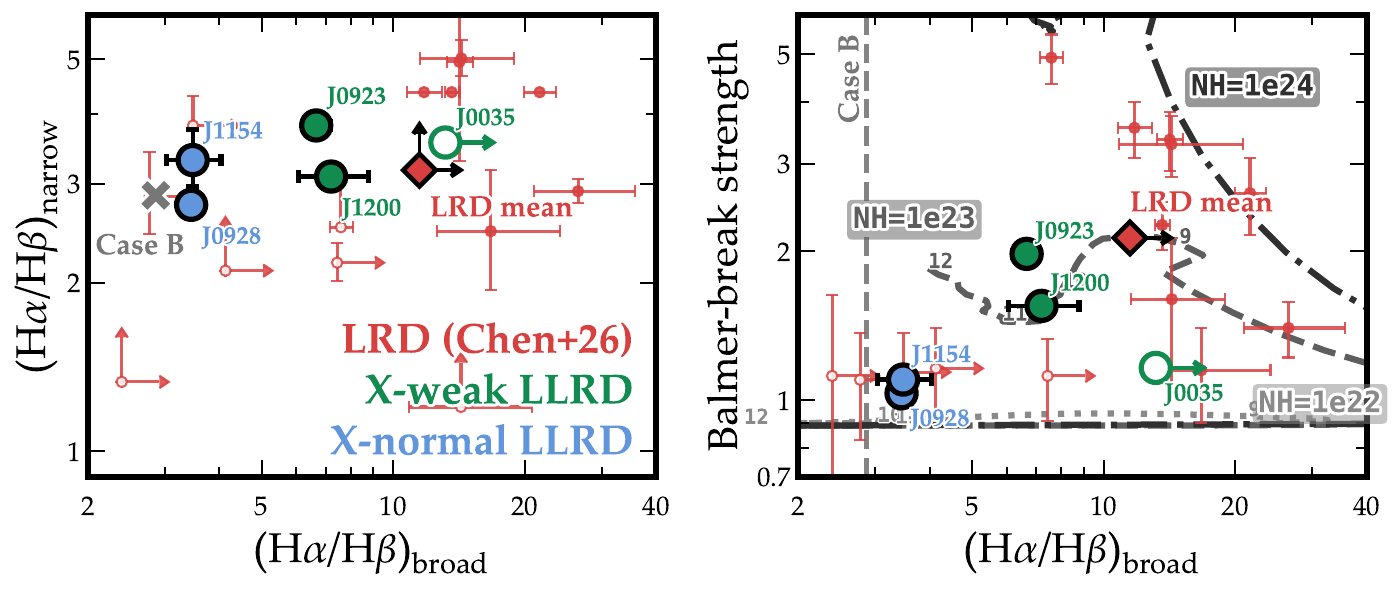}
    \caption{Balmer diagnostics support X-normal/weak division and moderate ($N_\mathrm{H}\sim10^{22-23}\,\mathrm{cm}^{-2}$) obscuration of X-weak LLRDs. Left: broad versus narrow \ha/\hb. Right: Balmer-break strength versus broad \ha/\hb. Red, green, and blue symbols denote LRDs from \citet{Chen2026Abcd}, X-weak LLRDs, and X-normal LLRDs, respectively. Arrows indicate limits, diamonds mark representative LRD average, and gray symbols indicate Case B. Curves in the right panel show the dense-gas calculation from \citet{Yan2026Balmer}: boxed labels identify $N_\mathrm{H}$ for each track, while numbers along the tracks give $n_\mathrm{H}$ in logarithm $\mathrm{cm}^{-3}$.}
    \label{fig:bdbb}
\end{figure*}

If X-weak LLRDs trace the same physical phenomena as high-redshift LRDs, as suggested by the multiwavelength similarities discussed in Section \ref{subsec:compare}, their proximity opens an X-ray regime otherwise inaccessible at high redshift. Indeed, with only $\sim$\sfrac{1}{1000} of the exposure time ($\sim\qty{40}{ks}$ versus $\sim\qty{40}{Ms}$ for the LRD stack of \citealt{Yue2024Stacking}), our observations provide tighter constraints than those available for $\qty{95}{\%}$ of the individual LRDs with measured $L_{2500}$ (Figure \ref{fig:lx_lha_luv}).

Figure \ref{fig:sed} compares the stacked IR--optical--UV--X-ray SEDs of the X-weak and X-normal LLRDs with the LRD SED from \citet{Akins2025Cosmosweb}. To emphasize differences in SED shape, all three are normalized at the rest-frame \qty{2500}{\AA} following convention in Figure \ref{fig:lx_lha_luv}. Remarkably, the X-weak LLRDs exhibit a prominent V-shaped optical--UV continuum closely resembling the LRD stack. The main difference occurs at rest-frame $\sim$\qtyrange[range-phrase=--,range-units=single]{2}{3}{\micro m}, where the X-weak LLRDs are detected at a luminosity comparable to the LRD upper limit. This difference may reflect a larger contribution from an evolved stellar population in the lower-redshift galaxies \citep[e.g.,][]{Bell2001Stellar,Bao2026Compact}.

The X-ray regime is where local analogs provide the strongest additional leverage. Figure \ref{fig:sed} compares the observed SEDs with representative unobscured super-Eddington \citep{Inayoshi2025Weakness}, CT, and Compton-thin models. The intrinsic X-ray normalizations of the absorption models are tied to the average X-weak LLRD UV luminosity using the standard unobscured-quasar relation of \citet{Chira2026Revisiting}.

We derive the LLRD X-ray SEDs from response-weighted broad-band measurements normalized by the stacked $L_{2500}$. The X-weak stack has a $\sim1.9\sigma$ soft-band excess, for which we adopt a $3\sigma$ upper limit, and a $3.3\sigma$ detection over \qtyrange[range-phrase=--,range-units=single]{2}{8}{keV}. Because \epfxt~effective area declines sharply above $\sim\qty{2.3}{keV}$, the latter is weighted toward rest-frame $\sim$\qtyrange[range-phrase=--,range-units=single]{2}{3}{keV} photons. By comparison, current Chandra constraints on high-redshift LRDs predominantly probe much harder rest-frame energies ($\gtrsim\qty{6}{keV}$). The X-weak LLRDs therefore extend the observational view of LRD-like systems down to rest-frame energies of $\sim\qty{1}{keV}$. The LLRD and LRD stacks differ by only $\lesssim\qty{0.4}{dex}$ in $L_{2500}$, corresponding to $\lesssim\qty{0.2}{dex}$ in expected normalized $L_\mathrm{2keV}$, making a normalized comparison possible.

The apparent hardness of the X-weak LLRD stack is in tension with the unabsorbed super-Eddington model shown in Figure \ref{fig:sed}. When normalized to the LLRD stack at \qty{2500}{\AA}, this very soft model, with $\Gamma\sim\num{3.4}$, underpredicts the measured \qtyrange[range-phrase=--,range-units=single]{2}{3}{keV} luminosity by $\sim\qty{1.5}{dex}$. At its native normalization, which is $\sim\qty{0.9}{dex}$ brighter at \qty{2500}{\AA}, the model remains $\sim\qty{0.6}{dex}$ below the \qtyrange[range-phrase=--,range-units=single]{2}{3}{keV} measurement while exceeding the soft-band upper limit by $\sim\qty{0.6}{dex}$. Thus, its spectral shape cannot reproduce both constraints, and we do not favor this particular unabsorbed super-Eddington model. 
Nevertheless, a super-Eddington scenario with stronger X-ray normalization \citep[e.g.,][]{Jin2017Supereddington}, modified by additional absorption \citep[e.g.,][]{Liu2025Balmer,Kido2025Black,Liu2026Optically,Madau2026LittleRedDots87d4e0bd} cannot be excluded with the present data. Observations with greater sensitivity above \qty{2}{keV}, for example with \xmm~and \nustar, will be required to distinguish this possibility from an intrinsically harder continuum.

Another widely discussed explanation for LRD X-ray weakness is CT obscuration, with $N_{\mathrm{H}}\gtrsim10^{24.5}\,\mathrm{cm}^{-2}$. Assuming the intrinsic X-ray luminosity expected from local AGN relations, the CT model with a local-AGN-like geometry \citep{Buchner2019Xray} is compatible with the very-hard-X-ray LRD upper limits but underpredicts the X-weak LLRD luminosity around \qtyrange[range-phrase=--,range-units=single]{2}{3}{keV} by more than \qty{2}{dex}. The measured X-weak LLRD $L_{2-10}^\mathrm{int}$ is $\sim\qty{27}{\%}$ of that predicted from the broad-\ha~luminosity using the \citet{Jin2012CombinedIIRelation} relation.

Additional soft X-ray emission scattered by a warm ionized medium could partially alleviate this tension. Such a component may be relevant given the strong \oiii~emission, with average EWs of $\sim$\qtyrange[range-phrase=--,range-units=single]{300}{400}{\AA}, observed in both LRDs \citep[e.g.,][]{deGraaff2025Little,Ji2026Compact} and LLRDs. However, the relation measured for local obscured AGNs by \citet{Gupta2021Spectroscopic} gives a conservative upper estimate of only $\sim5.6\pm1.5\%$ for the scattering fraction at the average LLRD $L_{\mathrm{[O\,III]}}/L_{\mathrm{2-10\,keV}}^{\mathrm{int}}$. Adopting a larger intrinsic $L_{\mathrm{2-10\,keV}}^\mathrm{int}$ (i.e., $N_\mathrm{H}>10^{22}\,\mathrm{cm}^{-2}$) would reduce this expected fraction even further. A separate possibility is that lower metallicity weakens photoelectric absorption, as potentially motivated by the dust-poor nature of LRDs \citep[e.g.,][]{Chen2025Dust,Setton2025Confirmed} and LLRDs. However, even in the limiting metal-free case, attenuation of the direct beam by Compton scattering alone gives a maximum transmission of $e^{-\tau_{\mathrm{T}}}=e^{-N_{\mathrm{H}}\sigma_{\mathrm{T}}}\simeq\qty{12}{\%}$ at $N_{\mathrm{H}}=10^{24.5}\,\mathrm{cm}^{-2}$. Combining this limiting direct transmission ($\qty{12}{\%}$) with the conservatively estimated scattered component ($\lesssim\qty{5}{\%}$) still falls below the observed value of $\qty{27}{\%}$. Thus, although the present data do not formally exclude CT obscuration, they are difficult to reconcile with the adopted CT model unless LLRDs have substantially stronger scattering, more prominent low-column-density leakage, or a different obscuring geometry from typical local obscured AGNs.

Importantly, the X-weak LLRDs show that a system satisfying the observational LRD criteria---with a compact morphology, broad \ha, and a V-shaped continuum---does not necessarily require extreme X-ray weakness.
Most individual high-redshift LRD limits remain too shallow to establish the magnitude of their X-ray deficits; indeed, \citet{Casey2026Population} showed that X-ray-detected local LRD-like quasars and ordinary quasars would frequently remain undetected under the published LRD limits. Our X-weak LLRD stack shows a moderate deficit of $\sim\qty{0.65}{dex}$ relative to the standard $L_{\mathrm{2keV}}^{\mathrm{int}}$--$L_{2500}$ relation, yet remains compatible with both the representative LRD stack and \qty{95}{\%} of the individual LRD upper limits. Current measurements therefore permit many LRDs to have similarly moderate deficits and do not require extreme X-ray weakness to be universal.
Such a moderate deficit is close to the conventional threshold used to classify X-ray-weak AGNs \citep[e.g.,][]{Ni2018Connecting} and is not surprising among local broad-line dwarf galaxies \citep[e.g.,][]{Dong2012Xray,Boorman2025Xmmnewton}, within a broader low-mass AGN population spanning a wide range of specific accretion rates \citep{Igo2026Incidence}. A Compton-thin absorber with $N_{\mathrm{H}}\sim10^{22-23}\,\mathrm{cm}^{-2}$, combined with a mildly weak corona with $\Delta\alpha_{\mathrm{ox}}\sim-0.1$ to $-0.2$, is sufficient to reproduce the X-weak LLRD measurements without invoking more extreme scenarios. Physically, this may arise from a less extended corona, so that a smaller fraction of the accretion power is dissipated in hot electrons and emitted as Comptonized X-rays \citep[e.g.,][]{Done2012Intrinsic,Kubota2018Physical,Hagen2024Systematic,Chen2025Average}.

The Balmer diagnostics provide an independent view of this absorbing medium. Figure \ref{fig:bdbb} compares the LLRDs with high-redshift LRDs from \citet{Chen2026Abcd} and the dense-gas calculations of \citet{Yan2026Balmer}. Despite lacking obvious Balmer absorption, the X-weak LLRDs show elevated broad-line decrements and stronger Balmer breaks in the same direction as the high-redshift LRDs, while the X-normal LLRDs lie closer to type 1 AGNs. Within the \citet{Yan2026Balmer} calculations, the X-weak LLRD average favors $n_\mathrm{H}\sim10^{9-11}\,\mathrm{cm}^{-3}$ and $N_\mathrm{H}\sim10^{22-23}\,\mathrm{cm}^{-2}$, independently consistent with the moderate obscuration favored by the stacked X-ray SED. J0923 lies between $N_\mathrm{H}\sim10^{23}\,\mathrm{cm}^{-2}$ and $10^{24}\,\mathrm{cm}^{-2}$ tracks while remaining undetected in X-rays, suggesting a comparatively larger column to be investigated in the future. Together, the progression from high-redshift LRDs through X-weak to X-normal LLRDs/type 1 AGNs motivates a possible sequence where the central engine becomes increasingly exposed. 

Finally, other local-analog searches provide a useful comparison. \citealt{Casey2026Population} selected LRD-like colors from SDSS low-redshift quasars, rather than starting with dwarf galaxies, and found that many such quasars would remain undetected under the published high-redshift LRD X-ray limits. Despite the different selections, their result and ours agree that LRD-like appearance and individual high-redshift X-ray non-detections do not, by themselves, require an extreme X-ray deficit. Our X-weak LLRDs additionally suggest a connection between X-ray weakness and the nuclear gas traced by the Balmer decrement and break. At a more X-ray faint extreme, the local analog known as \textit{the Egg} \citep{Lin2026Discovery} shows the prominent Balmer absorption and break and a much stronger X-ray deficit relative to ordinary AGNs \citep{Simmonds2016Some}. Its non-detection is compatible with CT obscuration if its intrinsic X-ray emission resembles that of a standard AGN \citep{Ji2026Lord}. These different local samples motivate testing whether Balmer diagnostics track X-ray emergence across the growing population of LRD local analogs \citep[e.g.,][]{Lin2026Lrds,Ding2026Discovery,Rinaldi2026Beyond}.


\section{Summary and future prospects}
Targeted \epfxt~observations of six LLRDs demonstrate how local analogs can bypass the high-redshift LRD X-ray bottleneck. Using $\sim\sfrac{1}{1000}$ of the exposure for current LRD stacks, we obtain constraints tighter than those available for \qty{95}{\%} of individual LRDs with measured $L_{2500}$ while directly probing rest-frame soft X-rays inaccessible to Chandra at $z\sim6$.

The observations reveal two X-ray classes. Two X-normal LLRDs exhibit normal to moderately enhanced X-ray emission, optical variability, and bluer optical continua. The X-weak subset, which most closely resembles high-redshift LRDs in optical/UV SED shape, Balmer diagnostics, and lack of detected variability, lies $\sim\qty{0.65}{dex}$ below the standard unobscured-quasar $L_{\mathrm{2keV}}^{\mathrm{int}}$--$L_{2500}$ relation. Stacking three X-weak LLRDs yields a $3.3\sigma$ detection in the rest-frame \qtyrange[range-phrase=--,range-units=single]{2}{8}{keV} band, driven mainly by \qtyrange[range-phrase=--,range-units=single]{2}{3}{keV} photons. The rest-frame \qtyrange[range-phrase=--,range-units=single]{0.5}{2}{keV} band is not significantly detected ($1.9\sigma$).
The moderate deficit measured from the X-weak LLRD stack remains compatible with current individual and stacked LRD constraints, showing that extreme X-ray weakness need not be universal.

The rest-frame soft-X-ray constraints are difficult to reconcile with either the typical unobscured super-Eddington model or the local-AGN-like CT geometry examined here, although absorbed super-Eddington spectra with stronger intrinsic X-ray emission, or CT configurations with substantially enhanced scattering or low-column-density leakage cannot be excluded. The simplest explanation consistent with both the X-ray and Balmer diagnostics is a mildly X-ray-weak corona, with $\Delta\alpha_{\mathrm{ox}}\sim-0.1$ to $-0.2$, viewed through Compton-thin material with $N_\mathrm{H}\sim10^{22-23}\,\mathrm{cm}^{-2}$. Deeper X-ray observations and growing samples of local LRD analogs will test whether this interpretation is representative and clarify the diversity and physical origin of LRD X-ray emission beyond what is currently feasible at high redshift.

\begin{acknowledgments}

This work was supported by the National Natural Science Foundation of China (grant nos. 12533006, 12192221, 12373016, 123B2042), and Guizhou Provincial Major Scientific and Technological Program XKBF (2025)010 and XKBF (2025)011. ZYZ acknowledges the support of the Shanghai Leading Talent Program of Eastern Talent Plan (LJ2025051) and the China-Chile Joint Research Fund 178GJHZ2025067MI. LCH was supported by the China Manned Space Program (CMS-CSST-2025-A09), the National Science Foundation of China (12233001), and the National Key R\&D Program of China (2022YFF0503401). ZI acknowledges support through the European Space Agency (ESA) Research Fellowship in Space Science.

We thank the referee and editors for their helpful comments that improved the manuscript. We thank Shouyi Wang for discussions on \efxtdas~and technical aspects of \epfxt, Shuying Zhou for scientific discussion during the CSST annual meeting. We also acknowledge \citet{Hviding2026Xray}, whose compilation we digitized to obtain the comparison data points from \citet{Yue2024Stacking}, \citet{Fu2026Discovery}, and \citet{Hviding2026Xray} shown in Figure~\ref{fig:lx_lha_luv}.

This work uses the GALEX UV Unique Source Catalog (GUVcat) AIS \citep{Morrissey2007Calibration,Bianchi2017Guvcat,Bianchi2020Galex,Bianchi2020Matched}, hosted by the Mikulski Archive for Space Telescopes (MAST). The GUVcat data used in this work are available at \dataset[https://doi.org/10.17909/T9-PYXY-KG53]{https://doi.org/10.17909/T9-PYXY-KG53}. MAST is hosted by the Space Telescope Science Institute, which is operated by the Association of Universities for Research in Astronomy, Inc., under NASA contract NAS 5--26555.

This work also uses the AllWISE Source Catalog \citep{Wright2010Wise,AllWISE2019Catalog} and SPHEREx Quick Release 2 spectral images \citep{Bock2026Spherex,Spherex2025Qr2}, hosted by the NASA/IPAC Infrared Science Archive (IRSA). IRSA is funded by the National Aeronautics and Space Administration and operated by the California Institute of Technology. This publication makes use of data products from the Wide-field Infrared Survey Explorer, a joint project of the University of California, Los Angeles, and the Jet Propulsion Laboratory/California Institute of Technology, funded by NASA. This publication also makes use of data products from SPHEREx, a joint project of the Jet Propulsion Laboratory and the California Institute of Technology, funded by NASA.

Funding for the Sloan Digital Sky Survey IV has been provided by the Alfred P. Sloan Foundation, the U.S. Department of Energy Office of Science, and the Participating Institutions. SDSS acknowledges support and resources from the Center for High-Performance Computing at the University of Utah. The SDSS website is \url{https://www.sdss4.org}.

SDSS is managed by the Astrophysical Research Consortium for the Participating Institutions of the SDSS Collaboration, including the Brazilian Participation Group, the Carnegie Institution for Science, Carnegie Mellon University, Center for Astrophysics $|$ Harvard \& Smithsonian (CfA), the Chilean Participation Group, the French Participation Group, Instituto de Astrof\'{i}sica de Canarias, The Johns Hopkins University, Kavli Institute for the Physics and Mathematics of the Universe (IPMU) / University of Tokyo, the Korean Participation Group, Lawrence Berkeley National Laboratory, Leibniz-Institut f\"{u}r Astrophysik Potsdam (AIP), Max-Planck-Institut f\"{u}r Astronomie (MPIA Heidelberg), Max-Planck-Institut f\"{u}r Astrophysik (MPA Garching), Max-Planck-Institut f\"{u}r Extraterrestrische Physik (MPE), National Astronomical Observatories of China, New Mexico State University, New York University, University of Notre Dame, Observat\'{o}rio Nacional / MCTI, The Ohio State University, Pennsylvania State University, Shanghai Astronomical Observatory, United Kingdom Participation Group, Universidad Nacional Aut\'{o}noma de M\'{e}xico, University of Arizona, University of Colorado Boulder, University of Oxford, University of Portsmouth, University of Utah, University of Virginia, University of Washington, University of Wisconsin, Vanderbilt University, and Yale University.

This work made use of data from LAMOST (the Large Sky Area Multi-Object Fiber Spectroscopic Telescope, also known as the Guoshoujing Telescope; \url{https://cstr.cn/31118.02.LAMOST}; \citealt{Cui2012Lamost}). LAMOST is a Chinese national mega-science facility operated by the National Astronomical Observatories, Chinese Academy of Sciences.

\end{acknowledgments}




%
\facilities{EP-FXT, GALEX, SDSS, SPHEREx, WISE}

\software{
HEAsoft (v6.28; HEASARC 2014),
\xspec~\citep{Arnaud1996Xspec},
\topcat~\citep{Taylor2005},
GNU Parallel Tool \citep{Tange2011a},
\fxtdas~v1.30 (\url{https://epfxt.ihep.ac.cn/analysis}), 
\efxtdas~(\url{https://github.com/AstroChensj/eFXTDAS}), 
\xstack~(\url{https://github.com/AstroChensj/Xstack})
}


\appendix

\section{LLRD \epfxt~observation summary}\label{app:obs_summary}
\begin{deluxetable*}{lccclcc}
\tabletypesize{\small}
\setlength{\tabcolsep}{5pt}
\tablecaption{Einstein Probe/FXT observation summary for the LLRDs\label{tab:lgd_obsids}}
\tablehead{
\colhead{Name} &
\colhead{R.A.} &
\colhead{Decl.} &
\colhead{ObsID} &
\colhead{Observing span} &
\colhead{Exposure} &
\colhead{Mode} \\
\colhead{} &
\colhead{(deg)} &
\colhead{(deg)} &
\colhead{} &
\colhead{(UTC)} &
\colhead{(ks)} &
\colhead{}
}
\startdata
\raisebox{1.35ex}{\textbf{J092834.85+292136.3}} & \raisebox{1.35ex}{142.1452} & \raisebox{1.35ex}{29.3601} & {\scriptsize \shortstack[l]{\rule{0pt}{2.2ex}11900576513, 11900576514 \\ 11900576517, 11900594305}} & \raisebox{1.35ex}{\mbox{2026-01-29--2026-02-07}} & \raisebox{1.35ex}{5.3} & \raisebox{1.35ex}{FF 01} \\
\tableline
\raisebox{2.35ex}{\textbf{J115438.96+065025.0}} & \raisebox{2.35ex}{178.6624} & \raisebox{2.35ex}{6.8403} & {\scriptsize \shortstack[l]{\rule{0pt}{2.2ex}11900528385, 11900545668 \\ 11900574849, 11900574850 \\ 11900574851}} & \raisebox{2.35ex}{\mbox{2025-12-26--2026-01-28}} & \raisebox{2.35ex}{7.2} & \raisebox{2.35ex}{FF 01} \\
\tableline
\raisebox{6.35ex}{\textbf{J120052.37+331238.4}} & \raisebox{6.35ex}{180.2182} & \raisebox{6.35ex}{33.2107} & {\scriptsize \shortstack[l]{\rule{0pt}{2.2ex}11900712450, 11900712451 \\ 11900712452, 11900712453 \\ 11900712454, 11900712455 \\ 11900712456, 11900712457 \\ 11900712458, 11900712459 \\ 11900719745, 11900719747 \\ 11900719749}} & \raisebox{6.35ex}{\mbox{2026-05-03--2026-05-10}} & \raisebox{6.35ex}{11.3} & \raisebox{6.35ex}{FF 01} \\
\tableline
\raisebox{5.35ex}{\textbf{J003515.16+084859.9}} & \raisebox{5.35ex}{8.8132} & \raisebox{5.35ex}{8.8167} & {\scriptsize \shortstack[l]{\rule{0pt}{2.2ex}11900458113, 11900458117 \\ 11900458118, 11900458119 \\ 11900465410, 11900465411 \\ 11900465412, 11900465414 \\ 11900465415, 11900512001 \\ 11900512002, 11900512003}} & \raisebox{5.35ex}{\mbox{2025-11-01--2025-12-13}} & \raisebox{5.35ex}{11.1} & \raisebox{5.35ex}{FF 01} \\
\tableline
\raisebox{6.35ex}{\textbf{J092359.89+415736.7}} & \raisebox{6.35ex}{140.9995} & \raisebox{6.35ex}{41.9602} & {\scriptsize \shortstack[l]{\rule{0pt}{2.2ex}11900610563, 11900610567 \\ 11900610569, 11900618244 \\ 11900618245, 11900618246 \\ 11900618497, 11900618498 \\ 11900618499, 11900618501 \\ 11900674177, 11900674178 \\ 11900674179}} & \raisebox{6.35ex}{\mbox{2026-02-19--2026-04-06}} & \raisebox{6.35ex}{15.7} & \raisebox{6.35ex}{FF 01} \\
\tableline
\raisebox{2.35ex}{\textbf{J085058.09+303053.6}} & \raisebox{2.35ex}{132.7421} & \raisebox{2.35ex}{30.5149} & {\scriptsize \shortstack[l]{\rule{0pt}{2.2ex}11900638728, 11900638729 \\ 11900638730, 11900638732 \\ 11900638733, 11900638734}} & \raisebox{2.35ex}{\mbox{2026-03-11--2026-03-12}} & \raisebox{2.35ex}{3.4} & \raisebox{2.35ex}{FF 01} \\
\enddata
\tablecomments{Dates are given in UTC. The exposure is the module-averaged value, $(t_{\mathrm{FXTA}}+t_{\mathrm{FXTB}})/2$, summed over the listed ObsIDs; it therefore represents the effective on-source exposure rather than the module-summed collecting time. ``FF'' and ``01'' denote the full-frame observing mode and thin-filter configuration, respectively.}
\end{deluxetable*}
\onecolumngrid
Table~\ref{tab:lgd_obsids} summarizes the pointed \epfxt~observations obtained through our program (PI: J.-X. Wang) for the six LLRDs analyzed in this work. The exposure times were planned to yield at least a $3\sigma$ detection for an intrinsic \qtyrange[range-phrase=--,range-units=single]{2}{10}{keV} luminosity \qty{0.5}{dex} below that expected from the standard broad-\ha~AGN relation. J0850 was assigned a shorter exposure because its broad-\ha~luminosity is approximately \qty{1}{dex} lower than those of the other targets; for this source, the observational goal was to improve upon the archival \erosita~upper limit rather than to secure an individual detection.

\section{\efxtdas: enhanced data analysis tools for EP-FXT}\label{app:efxtdas}
\begin{figure*}
    \centering
    \includegraphics[width=0.9\linewidth]{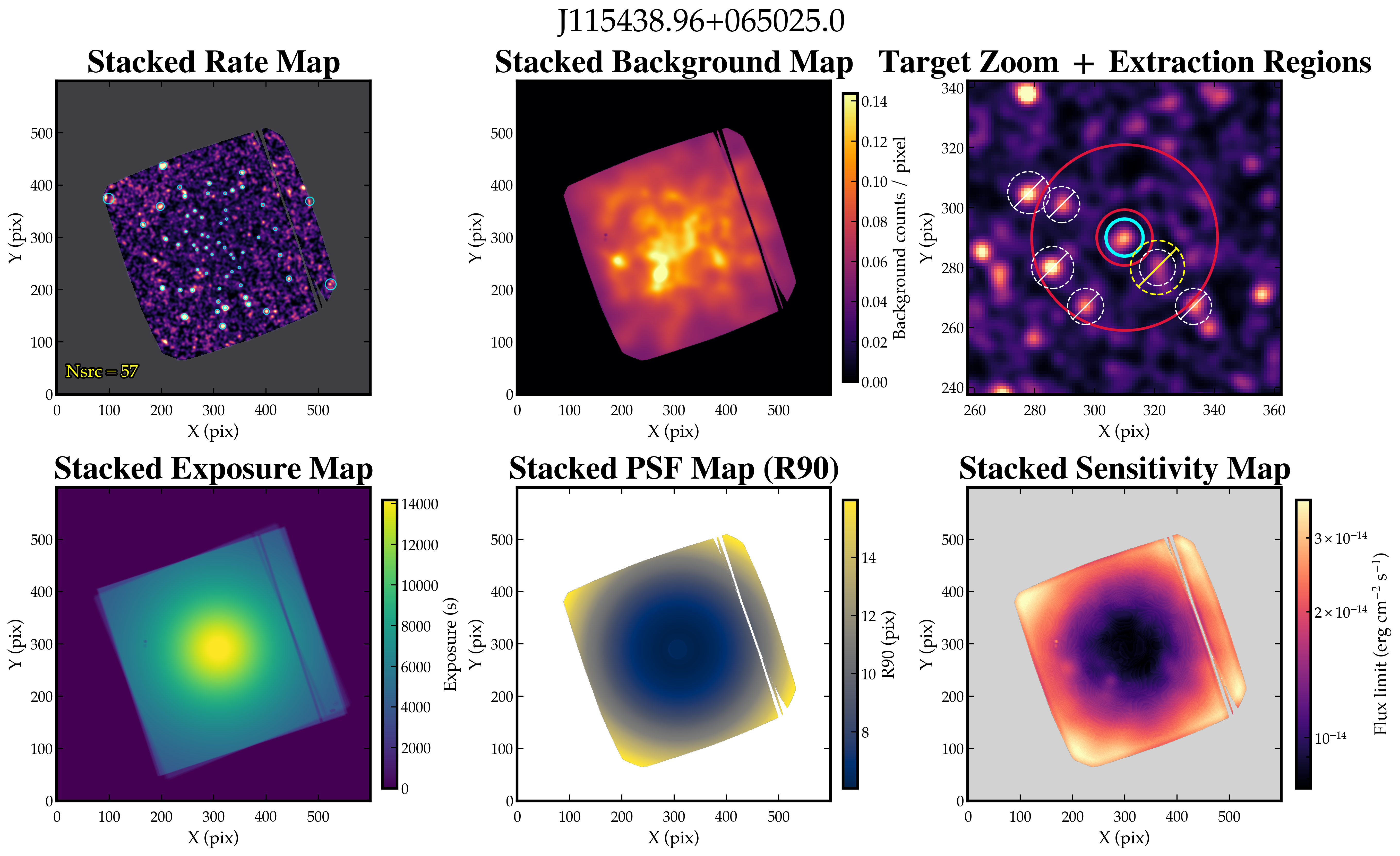}
    \caption{Example \efxtdas~quality-assessment summary for J115438.96+065025.0, generated with \texttt{fxtcombine-quickview}. From upper left to lower right, the panels show the stacked count-rate map and detected-source catalog, modeled background map, target-centered extraction geometry (overlaid with contaminating sources in the source/background extraction regions), stacked exposure map, 90\%-encircled-energy PSF-radius map, and limiting-flux sensitivity map.}
    \label{fig:efxtdas_quickview}
\end{figure*}

The official \fxtdas~software (version 1.30; CALDB version 1.30) provides the basic calibration and event-cleaning workflow for individual \epfxt~observations. Scientific analysis nevertheless requires additional capabilities, including source detection, contaminant-aware region construction, response generation for non-trivial extraction geometries, and the combination of imaging and spectral products from multiple observations and detector modules. We developed \efxtdas~to provide these capabilities and orchestrate the official tasks within a reproducible end-to-end workflow. The software is publicly available at \url{https://github.com/AstroChensj/eFXTDAS}; its top-level task, \texttt{fxtcombine}, performs the following steps.



\begin{enumerate}

\item \textit{Calibration and preparation of individual observations.} During the first iteration over the observations, \texttt{fxtcombine} processes each OBSID and FXT module separately. It invokes the official \fxtdas~tasks for coordinate reconstruction, energy calibration, particle identification, bad-pixel filtering, event grading, and construction of the basic good-time intervals. For observations obtained in full-frame mode, the corresponding \texttt{fsaevt} data are used to identify periods of enhanced instrumental background. The \texttt{fxtbkgoptrate} task derives an optimized background-rate threshold from the \texttt{fsaevt} light curve (defaults to \texttt{robust\_iqr.}), and the resulting flare-filtered good-time intervals are applied consistently to both the \texttt{fsaevt} and science event files.

The cleaned science events are then used to extract the \qtyrange[range-phrase=--,range-units=single]{0.5}{4}{keV}\footnote{This energy range is chosen for visualization and later source detection, because it maximizes the source-to-background ratio: the \epfxt~effective area declines while the instrumental background becomes increasingly important at higher energies. By contrast, the extraction-based significance quoted in main text and Appendix \ref{app:detection_significance} is evaluated over \qtyrange[range-phrase=--,range-units=single]{0.5}{8}{keV}, retaining the full calibrated spectral band.} counts image and to generate the corresponding vignetting-corrected exposure map. For each observation, \texttt{fxtpsfgen} also evaluates the calibrated energy- and off-axis-angle-dependent PSF and stores it as an observation PSF product containing the local encircled-energy-fraction curves. These per-observation images, exposure maps, and PSF products form the inputs to the subsequent stacking analysis.

\item \textit{Image and PSF stacking.} For each target, the cleaned events from all OBSIDs and both FXT modules are reprojected onto a common celestial frame and accumulated to form the stacked counts image. The corresponding exposure maps are reprojected and summed on the same grid. An exposure-corrected count-rate map is then obtained by dividing the stacked counts map by the stacked exposure map; this rate map is used primarily for visualization.

The effective PSF of the stacked observation is constructed from the individual observation PSF products. At each position in the stacked image, the local encircled-energy curves from the contributing observations are averaged using their reprojected exposure values as weights. The radius enclosing a requested fraction of the source counts, such as the 90\% encircled-energy radius, is then obtained by inverting this exposure-weighted stacked curve. This procedure accounts for the different off-axis angles and exposure contributions of the individual observations.

\item \textit{Source detection.} Source detection is performed with \texttt{fxtsrcdet} on the stacked \qtyrange[range-phrase=--,range-units=single]{0.5}{4}{keV} counts image, using the stacked exposure map and PSF product. A multiscale Mexican-hat wavelet search first identifies provisional source candidates. Significant candidate regions are temporarily carved from the image, and the remaining counts are adaptively smoothed relative to the exposure map to construct the background model. The candidate list is then refined through PSF-based Cash-statistic fitting. Sources sufficiently close for their PSFs to overlap are fitted simultaneously to reduce source confusion and flux cross-contamination.

For a point-source candidate, \texttt{fxtsrcdet} compares the best-fitting source-plus-background model with the background-only model and defines
\begin{equation}
    \texttt{DETLIKE}=-\ln p,\qquad
    p=Q\left(\frac{3}{2},\frac{\Delta C}{2}\right),
    \label{eq:efxtdas_detlike}
\end{equation}
where $\Delta C$ is the improvement in the Cash statistic and $Q$ is the regularized upper incomplete gamma function\footnote{See also \url{https://erosita.mpe.mpg.de/dr1/eSASS4DR1/eSASS4DR1_tasks/ersensmap_doc.html}.}. The three effective degrees of freedom correspond to the source amplitude and two centroid coordinates. Candidates with $\texttt{DETLIKE}>6$ are retained in the clean source catalog, following a conventional threshold used in \erosita-like source detection \citep{Liu2022Erosita}\footnote{As per \erosita~simulations \citep{Seppi2022Detecting}, this threshold corresponds roughly to \qty{14}{\%} false positive rate.}. This image-fitting likelihood is distinct from the extraction-based SNR defined in Equation~\ref{eq:det_snr}.

\item \textit{Construction of extraction regions.} 
The \texttt{fxtregions} task matches the source catalog to the optical target position within $\ang{;;8.6}$, corresponding approximately to the \qty{90}{\%}-confidence \fxt~positional accuracy, including systematic uncertainty \citep{Center2024Einstein}. When a counterpart is identified, the fitted X-ray centroid is adopted as the extraction center; otherwise, the optical position is used.

For the present analysis, \texttt{fxtregions} creates a circular source region of radius $\ang{;;60}$, enclosing $\sim\qty{90}{\%}$ of the on-axis PSF energy, and a background annulus spanning $\ang{;;90}$--$\ang{;;300}$. Neighboring sources are excised from the source and background regions with circular masks whose radii are determined from their fitted PSF profiles. Within the background annulus, each mask extends to the radius at which the contaminant surface brightness falls below 10\% of the local background. If a contaminant overlaps the source aperture, its mask instead extends to the radius at which its surface brightness falls below 20\% of the modeled target surface brightness. For an undetected target, for which the latter quantity cannot be constrained, the background-based criterion is used instead.

\item \textit{Spectral extraction and response generation.} After the regions have been established from the stacked image, \texttt{fxtcombine} performs a second iteration over all OBSIDs and detector modules. Source and background spectra are extracted separately from each cleaned event file using the same sky regions. The corresponding ARF and RMF are generated with \texttt{fxtrspgen}.

The standard \fxtdas~\texttt{fxtarfgen} task does not account for a compound source aperture consisting of a positive region with one or more subtracted contaminant regions. Such a source-source collision case should be common especially in deep exposures, given the relative broad PSF of \epfxt. In our custom \texttt{fxtrspgen} task, we rasterize the complete DS9 region expression and calculates the response for the retained aperture. The extracted ARF includes the spatially weighted vignetting correction, the energy-dependent PSF fraction enclosed by the aperture, and the fractional region coverage on the exposure map. The response therefore remains consistent with the source spectrum even when part of the nominal circular aperture has been removed.

\item \textit{Spectral stacking and diagnostic products.} Finally, the source and background PI spectra and their associated responses from the individual observations and detector modules are combined in observed-frame using \xstack~in its \texttt{same-target} mode. The source and background counts are summed directly, while the full response matrices ($\mathrm{ARF}\times\mathrm{RMF}$) are combined with exposure weighting and normalized by the total exposure before being decomposed into the stacked ARF and RMF. The stacked \texttt{EXPOSURE} is the sum of the individual exposures, whereas \texttt{AREASCAL}, \texttt{BACKSCAL}, and \texttt{CORRSCAL} are assigned their respective arithmetic means, separately for the source and background spectra. This produces a single stacked source spectrum and associated background and response files for each target, with a response representing the exposure-weighted contributions of the individual observations.

For the convenience of users, \texttt{fxtcombine} also generates a sensitivity map from the stacked background, exposure, and PSF products. At each image position, the task calculates the limiting source flux for the adopted encircled-energy aperture and detection-likelihood threshold. The complete set of imaging and diagnostic products is collected by \texttt{fxtcombine-quickview} into the summary shown in Figure~\ref{fig:efxtdas_quickview}. This combined view allows the stacked rate, background, exposure, PSF, sensitivity, and extraction-region products to be inspected together before the final spectra are analyzed.

\end{enumerate}

\section{Extraction-based detection significance}\label{app:detection_significance}

For a chosen energy band, we sum the source-aperture counts to obtain $S$ and the background-spectrum counts to obtain $B$. The background uncertainty $\sigma_B$ is obtained by adding the channel-level \texttt{STAT\_ERR} values in quadrature. The background-to-source scaling factor is
\begin{equation}
    \alpha =
    \frac{\texttt{EXPOSURE}_\mathrm{src}}{\texttt{EXPOSURE}_\mathrm{bkg}}
    \frac{\texttt{BACKSCAL}_\mathrm{src}}
         {\texttt{BACKSCAL}_\mathrm{bkg}}
    \frac{\texttt{AREASCAL}_\mathrm{src}}
         {\texttt{AREASCAL}_\mathrm{bkg}},
    \label{eq:det_alpha}
\end{equation}
where \texttt{EXPOSURE}, \texttt{BACKSCAL}, \texttt{AREASCAL} are standard OGIP keywords.

Under the background-only hypothesis, the expected source-aperture counts are $\alpha b$, where $b\geq0$ is the unknown true background level corresponding to the background spectrum. We treat $B\pm\sigma_B$ as a Gaussian measurement of $b$ and adopt a flat prior over $b\geq0$. The marginalized background-only probability is then
\begin{equation}
P_{\rm Pois}(N\geq S\mid\alpha b) =
\sum_{n=S}^{\infty}
\frac{\exp(-\alpha b)(\alpha b)^n}{n!},
\label{eq:det_poisson_tail}
\end{equation}
and
\begin{equation}
p_{\rm null} =
\frac{
\displaystyle \int_0^\infty
P_{\rm Pois}(N\geq S\mid\alpha b)
\exp\left[-\frac{(b-B)^2}{2\sigma_B^2}\right] \,{\rm d}b
}{
\displaystyle \int_0^\infty
\exp\left[-\frac{(b-B)^2}{2\sigma_B^2}\right] \,{\rm d}b
}.
    \label{eq:det_pnull}
\end{equation}

We convert this probability to a one-sided Gaussian-equivalent significance,
\begin{equation}
    {\rm SNR} =
    \max\left[0,\Phi^{-1}(1-p_{\rm null})\right],
    \label{eq:det_snr}
\end{equation}
where $\Phi$ is the standard-normal cumulative distribution function. Values with $p_{\rm null}>0.5$ are reported as zero significance. This aperture-based statistic is distinct from the image-based \texttt{DETLIKE}, which is derived from PSF fitting.

\section{Balmer decrement/break measurements}\label{app:balmer_decrement}
\begin{deluxetable*}{ccccccc}
\tabletypesize{\footnotesize}
\tablewidth{\textwidth}
\setlength{\tabcolsep}{6pt}
\renewcommand{\arraystretch}{1.22}
\tablecaption{Balmer-line ratios and Balmer-break strengths of the LLRDs\label{tab:llrd_balmer}}
\tablehead{
\colhead{X-ray class} &
\colhead{Name} &
\multicolumn{2}{c}{$\mathrm{H}\alpha/\mathrm{H}\beta$} &
\multicolumn{2}{c}{$\mathrm{H}\gamma/\mathrm{H}\alpha$} &
\colhead{Balmer-break strength} \\
\colhead{} &
\colhead{} &
\colhead{Narrow} &
\colhead{Broad} &
\colhead{Narrow} &
\colhead{Broad} &
\colhead{}
}
\startdata
X-normal & J0928 & $2.75_{-0.10}^{+0.10}$ & $3.45_{-0.10}^{+0.11}$ & $0.161_{-0.008}^{+0.008}$ & $0.117_{-0.009}^{+0.009}$ & $1.033\pm0.006$ \\
X-normal & J1154 & $3.31_{-0.35}^{+0.44}$ & $3.47_{-0.44}^{+0.56}$ & $0.156_{-0.037}^{+0.037}$ & $<0.276$ & $1.101\pm0.048$ \\[2pt]
\tableline
\noalign{\vskip 2pt}
X-weak & J1200 & $3.09_{-0.07}^{+0.08}$ & $7.21_{-1.14}^{+1.58}$ & $0.141_{-0.005}^{+0.005}$ & $<0.169$ & $1.551\pm0.016$ \\
X-weak & J0035 & $3.55_{-0.06}^{+0.06}$ & $>13.16$ & $0.120_{-0.003}^{+0.003}$ & $<0.035$ & $1.163\pm0.024$ \\
X-weak & J0923 & $3.81_{-0.15}^{+0.16}$ & $6.66_{-0.46}^{+0.51}$ & $0.117_{-0.009}^{+0.009}$ & $0.068_{-0.010}^{+0.011}$ & $1.976\pm0.046$ \\
X-weak & J0850 & $2.44_{-0.13}^{+0.15}$ & -- & $0.271_{-0.022}^{+0.022}$ & -- & $0.794\pm0.042$ \\[2pt]
\enddata
\tablecomments{The X-ray classes follow the grouping adopted in the main analysis.  All ratios and break strengths are corrected for Galactic extinction but not for attenuation internal to the sources.  Errors on measured line ratios are 16th--84th percentile intervals from the joint posterior; ``$<$'' and ``$>$'' denote one-sided Gaussian-equivalent $3\sigma$ limits (the 99.865th or 0.135th posterior percentile).  Dashes mark undefined broad ratios when broad H$\alpha$ itself is not significantly detected. Break errors are propagated statistical errors. The break is defined by Equation~\ref{eq:balmer_break}.}
\end{deluxetable*}
\onecolumngrid
We measured the Balmer-line ratios uniformly from the original SDSS/LAMOST spectra rather than combining the independently fitted catalog fluxes from \citet{Lin2025Discovery}. After correcting each rest-frame spectrum for Galactic extinction with the SFD reddening map and an $R_V=3.1$ law \citep{Fitzpatrick1999Correcting}, we determined the stellar continuum and Balmer absorption with \ppxf~and E-MILES templates \citep{Cappellari2017Improving}. The Balmer emission regions were masked during this step. \ppxf~freely determined the mixture of stellar templates and their velocity shift and broadening, while an additive polynomial absorbed smooth continuum-shape and calibration residuals. We repeated the continuum fit over a modest range of polynomial orders, adopted the median prediction, and included their scatter in the line-spectrum uncertainty. The adopted stellar-continuum model was then subtracted from the spectrum before the emission-line analysis, in which only a local linear residual continuum was allowed to vary within each fitting window.

We jointly fitted the \hg~(\qtyrange[range-phrase=--,range-units=single]{4260}{4430}{\AA}), \hb~(\qtyrange[range-phrase=--,range-units=single]{4750}{5105}{\AA}), and \ha~(\qtyrange[range-phrase=--,range-units=single]{6450}{6800}{\AA}) windows. The six narrow and broad Balmer fluxes were independently free and nonnegative. Given the limited spectral quality, the three narrow lines shared one free velocity and width, and the three broad lines shared a separate free velocity and width. The fit also included the neighboring \oiii, \nii, and \sii~emission. Their principal amplitudes were free, while the \oiii~and \nii~doublet ratios were fixed to their theoretical values. Narrow neighboring lines shared the narrow Balmer kinematics. Additional wide \oiii~and \ha~components had free amplitudes and fixed kinematics. All profiles included instrumental broadening assuming resolving power $R=1800$.

Line ratios are evaluated for each joint-posterior sample, thereby retaining covariance between the shared profiles. We report posterior medians and 16th--84th percentile intervals for detections, and one-sided $3\sigma$ limits otherwise (Table \ref{tab:llrd_balmer}). J0923 has $({\rm H}\alpha/{\rm H}\beta)_{\rm broad}=6.66_{-0.46}^{+0.51}$ and $({\rm H}\gamma/{\rm H}\alpha)_{\rm broad}=0.068_{-0.010}^{+0.011}$. For J0035, the nondetections of broad H$\beta$ and H$\gamma$ instead give $({\rm H}\alpha/{\rm H}\beta)_{\rm broad}>13.16$ and $({\rm H}\gamma/{\rm H}\alpha)_{\rm broad}<0.035$.

Independently of the line fit, we measured the Balmer break directly from the original, continuum-included spectrum. No \ppxf~continuum model or emission-line model enters this measurement. We define
\begin{equation}
 \mathrm{Balmer\text{-}break\ strength}\equiv
 \frac{\left\langle f_\nu\right\rangle_{4000-4500\,\mathring{\rm A}}}
      {\left\langle f_\nu\right\rangle_{3000-3500\,\mathring{\rm A}}},
 \label{eq:balmer_break}
\end{equation}
where each mean is a wavelength-integrated mean flux density per unit frequency. J0923 has the strongest local break, $1.976\pm0.046$, whereas J0035 has only $1.163\pm0.024$ despite its more extreme line-ratio limits. The two diagnostics therefore need not vary together source by source. Interestingly, J0923 also has a radio counterpart with reported variability \citep{An2026Radio}, potentially tracing jet- or wind-driven shocks in its gas-rich nuclear environment. A detailed joint radio/X-ray investigation is deferred to future work.


\section{Spectral-fitting details and the X-normal LLRDs}\label{app:Xnormal}

The spectral models and stacking procedure are described in Section \ref{subsec:compare}. Here we provide additional fitting details and present the spectra of the two X-normal LLRDs.

Individual spectra are grouped to contain at least five background counts per bin and fitted over the observed-frame \qtyrange[range-phrase=--,range-units=single]{0.5}{8}{keV} band using the C-statistic \citep{Buchner2023Statistical}. The higher-count stacked spectra are fitted over the rest-frame \qtyrange[range-phrase=--,range-units=single]{0.5}{8}{keV} band using the PG-statistic \citep{Chen2025Average}. We use $\Delta C=1$ to determine the quoted $1\sigma$ luminosity and parameter uncertainties for detected sources, and $\Delta C=9$ to derive the $3\sigma$ luminosity upper limits for nondetections.

The fixed host component in Section \ref{subsec:compare} is \texttt{zTBabs}\textsubscript{\texttt{host}}\texttt{*(zpowerlw}\textsubscript{\texttt{xrb}}\texttt{+mekal)}. We adopt $N_\mathrm{H,host}=10^{21}\,\mathrm{cm}^{-2}$ and $\Gamma_\mathrm{xrb}=1.8$, and normalize the X-ray-binary component from the stellar mass and SFR following \citet{Lehmer2010Chandra}. The \mekal~component has $kT=\qty{0.24}{keV}$ and is normalized from the SFR following \citet{Mineo2012Xray}. As noted in the main text, these host components are negligible at the present sensitivity.

\begin{figure}
    \centering
    \includegraphics[width=0.7\linewidth]{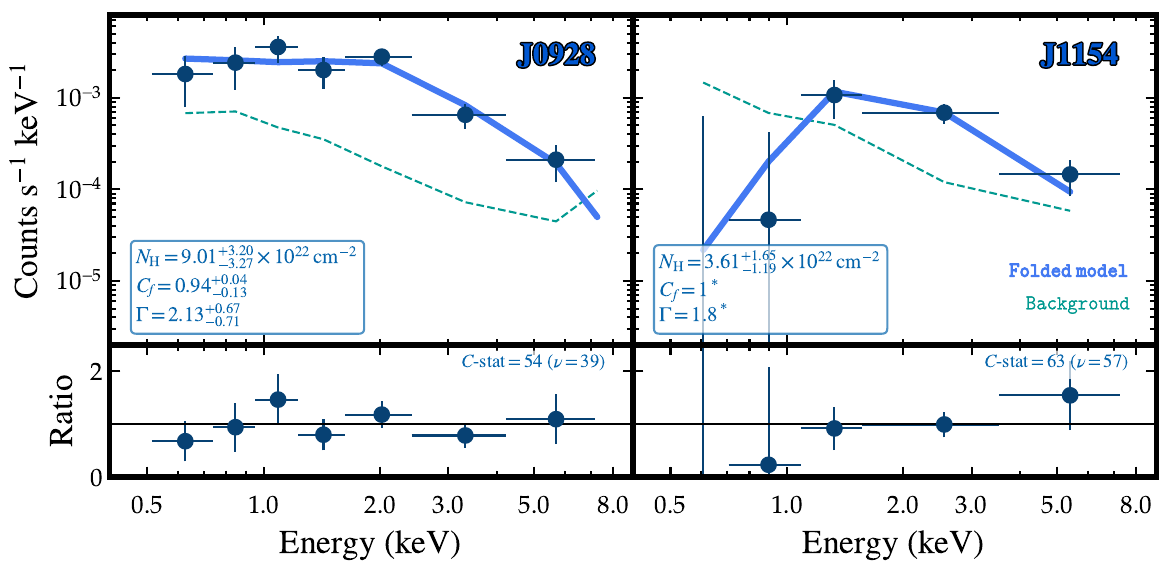}
    \caption{Observed-frame
    \qtyrange[range-phrase=--,range-units=single]{0.5}{8}{keV} \epfxt~spectra of the X-normal LLRDs J0928 (left) and J1154 (right). Upper panels show the spectra, response-folded best-fitting models, and expected backgrounds; lower panels show the data-to-model ratios.}
    \label{fig:xray_spec}
\end{figure}

For J1154, the photon index is fixed at $\Gamma=1.8$, while the intrinsic column density and power-law normalization are free. Its absorption-corrected X-ray luminosity is consistent with the standard type~1 AGN relations shown in Figure~\ref{fig:lx_lha_luv}.

J0928 shows a prominent soft residual when fitted with the baseline model and $\Gamma=1.8$. Allowing $\Gamma$ to vary improves the fit by $\Delta C=9.45$ but yields an unusually flat continuum, $\Gamma\sim0.5$, with negligible intrinsic absorption. Replacing the fully covering absorber with \texttt{zpcfabs} improves the fit by a further $\Delta C=4.73$ and gives $N_\mathrm{H}=9.01_{-3.27}^{+3.20}\times10^{22}\,\mathrm{cm}^{-2}$, $C_f=0.94_{-0.13}^{+0.04}$, and $\Gamma=2.13_{-0.71}^{+0.67}$. J0928 also shows moderately enhanced intrinsic X-ray emission relative to the standard AGN relations.

The simplest interpretation of the two X-normal LLRDs is that they are relatively normal type~1 AGNs hosted by metal-poor dwarf galaxies \citep[from the BPT analysis of][]{Lin2025Discovery}. A more speculative possibility is that they represent a later evolutionary stage in which an initially LRD-like nucleus is becoming exposed \citep[e.g.,][]{Fu2026Discovery,Hviding2026Xray}. This interpretation is motivated by the partial-covering absorber and enhanced X-ray emission of J0928, together with the bluer optical continuum and detected variability of the X-normal LLRDs \citep{Lin2026Local}. In this picture, radiative feedback may have begun to clear the surrounding material, revealing both the central X-ray source and the blue accretion-disk continuum. Further multiwavelength observations are needed to test this evolutionary interpretation.

\section{Kaplan-Meier estimate of the X-ray luminosity distribution}\label{app:km}
Treating the three undetected sources as left-censored at their tabulated $3\sigma$ $\log L_{2-10}^\mathrm{int}$ limits, we computed a reverse Kaplan–Meier estimate \citep{Kaplan1958Nonparametric} for the six individual LLRDs. Its cumulative probability reaches $2/3$, $5/6$, and $1$ at $\log L_{2-10}^\mathrm{int}=42.80$, $43.48$, and $44.09$, respectively; the lowest third remains unresolved below the detections. With only three detections, and with limits and intrinsic luminosities dependent on the adopted spectral model, we use this estimate descriptively rather than to define the X-normal/X-weak groups.


\bibliography{lgd_xray}{}
\bibliographystyle{aasjournalv7}



\end{document}